\documentclass[aps,prl,reprint,superscriptaddress, amsmath, amssymb,floatfix]{revtex4-2}
\usepackage{graphicx}
\usepackage{dcolumn}
\usepackage{bm}
\newlength{\myparskip}
\usepackage[utf8]{inputenc}
\usepackage{tikz}
\usetikzlibrary{plotmarks}
\usetikzlibrary{shapes}

\usepackage{multirow}
\begin{document}
{
  \title{\bf Jammed solids with pins: \\ Thresholds, Force networks and Elasticity} 
  
  \author{Andy L. Zhang} 
  \affiliation{Department of Physics and Astronomy, Swarthmore College}
  \author{Sean A. Ridout}
  \affiliation{Department of Physics and Astronomy, University of Pennsylvania}
  \author{Celia Parts}
\affiliation{Department of Physics and Astronomy, Swarthmore College}
  \author{Aarushi Sachdeva}
  \affiliation{Department of Physics and Astronomy, Swarthmore College}
    \author{Cacey S. Bester}
\affiliation{Department of Physics and Astronomy, Swarthmore College}
 \author{Katharina Vollmayr-Lee}
   \affiliation{Department of Physics and Astronomy, Bucknell University}
 \author{Brian C. Utter}
\affiliation{ Department of Physics, University of California at Merced} 
\author{Ted Brzinski}
\affiliation{Department of Physics and Astronomy, Haverford College}
\author{Amy L. Graves}
\affiliation {Department of Physics and Astronomy, Swarthmore College}
\date{\today}

\begin{abstract}
{\em Abstract:} The role of fixed degrees of freedom in soft/granular matter systems has broad applicability and theoretical interest.  Here we address questions of the geometrical role that a scaffolding of fixed particles plays in tuning the threshold volume fraction and force network in the vicinity of jamming. Our 2d simulated system consists of soft particles and fixed ``pins", both of which harmonically repel overlaps. On one hand, we find that many of the critical scalings associated with jamming in the absence of pins continue to hold in the presence of even dense pin latices. On the other hand, the presence of pins lowers the jamming threshold, in a universal way at low pin densities and a geometry-dependent manner at high pin densities, producing packings with lower densities and fewer contacts between particles.  The onset of strong lattice dependence coincides with the development of bond-orientational order. Furthermore, the presence of pins dramatically modifies the network of forces, with both unusually weak and unusually strong forces becoming more abundant. The spatial organization of this force network depends on pin geometry and is described in detail. Using persistent homology we demonstrate that pins modify the topology of the network.  Finally, we observe clear signatures of this developing bond-orientational order and broad force distribution in the elastic moduli which characterize the linear response of these packings to strain. 

\end{abstract}
  \maketitle

}
\vspace{1cm}
\section{Introduction}
\label{sec:intro}
Jamming of soft or granular materials constitutes a transition from fluid-like to a solid-like state that can support finite stress.  The disordered geometry of grain-grain contact forces is a key aspect of these solids \cite{AndreottiCh3}.The out-of-equilibrium nature of jammed phases has supported much activity in development of new statistical models \cite{Makse2010, Goodrich2016}.  Traditionally, the transition to jamming is marked by control parameters of system density, applied stress, and the analog of temperature for macroscopically large grains; and is depicted in the Liu-Nagel jamming phase diagram \cite{Liu1998, Ohern2003}.   To this venerable diagram, it has been suggested that a new axis might be added, representing the density of quenched disorder \cite{Reichhardt2012}. This indicates an interest in the jamming and glass communities on how quenched disorder in the form of pinned grains affects the transition of systems, either under applied drive, or in its absence: the so-called ``Point J".

One might be motivated to study matter with fixed degrees of freedom for many reasons.  Flowing states of matter are impacted by obstacle-filled substrates in both ordered and random geometries \cite{Stoop2020, Reichhardt2021}.  Pinned inclusions have modified phase transition behavior \cite{Deutschlander2013, Qi2015} and obstacle lattices have been used to sort biomaterials \cite{Chou1999, Chakrabarti2020}. Pinned particles in glass-formers can tune spatial heterogeneity, kinetic fragility and the transition point \cite{Cammarota2012, Berthier2012, Ozawa2015, Angelani2018}. Scaling theory near Point J can be extended to describe quenched disorder via a pinning susceptibility \cite{Graves2016, Das2017}.

 So called “partly pinned systems” have served as a theoretically-advantageous model, in which a fraction of equilibrated, fluid particles are pinned to serve as a confining matrix, through which the remaining, mobile particles flow \cite{Krakoviack2010}. In the current study, in contrast with simulations where particle positions are frozen during the creation of a fluid, jammed or glassy state, pins are placed at the outset of the simulation, and are centers of force of negligible size. Their role is both to exclude volume and to scaffold the emerging jammed structure.  We find that pins produce some surprises, in terms of how they tune the jamming threshold, reduce mean contact number, and modify the network of forces - enhancing the likelihood both of weak and strong forces at jamming. These changes are reflected in the linear elastic properties of the solid.

\section{Simulation details}
\label{sec:simulation}


Particles are frictionless, soft, repulsive discs: the so-called ``Ising Model" of jamming for their simplicity, yet fidelity in reproducing the physics of more realistic models. $N$ frictionless particles and $N_f$ dramatically smaller, fixed particles called ``pins" interact via the well-studied harmonic, repulsive potential:
\begin{align} V = \left\{
\begin{matrix}
0 & r_{ij}>d_{ij}\\
\epsilon(1-\frac{r_{ij}}{d_{ij}})^2 & \mbox{otherwise}
\end{matrix} 
\right.  \ \ ,
\label{eq:potential}
\end{align}
where $r_{ij}$ is the distance between the centers of particles $i$ and $j$, and $d_{ij}$ is the sum of particle radii.
Equal numbers of large and small particles with the size ratio $r_L/r_S$ = 1.4, known to discourage ordered packing, are initially placed at random in a two-dimensional simulation cell with periodic boundary conditions.  Pin radii are roughly a factor of $1000$ smaller than those of small particles; results are independent of this size ratio.  We study numbers of particles in the range $N = 230-920$. $N_f$ particles (a number which varies depending on the type of analysis performed) are fixed in a desired lattice geometry: square, triangular, honeycomb or random.

We will see in what follows that both pin density for a given lattice geometry, and choice of lattice geometry serve as tuning parameters of the packing. One can characterize the pin density by introducing the parameter $\alpha$, the ratio of area of a particle to area of box per pin.  However, given pins are placed in different Bravais lattices with lattice constant $a$, it is also worthwhile to characterize the density of pins by the ratio of length scales $\lambda \equiv r_S/a$. 
These are related via 
\begin{equation}
\lambda^2 = \alpha \ \frac{g}{c}
\label{eq:lambda_alpha}
\end{equation}
where $g$ is a geometrical factor equal to $1, \ \sqrt{3}/2, \  3 \sqrt{3}/4$ for square, triangular and honeycomb lattices and $c = \frac{\pi}{2} ( 1 + 1.4^2)$.  
A third measure of pin density which will be convenient is $N_f/N$. When curves from multiple pin densities are shown on the same plot, each color will represent a fixed value of $N_f/N$ so that square and triangular lattices may use the same color scheme.  It is the case that:
\begin{equation}
\alpha = \phi \ N_f/N \ \ .
\label{eq:NfOverN}
\end{equation}

\vspace{3 mm}

Energy is minimized via the FIRE algorithm \cite{FIRE}.
Final configurations will be considered jammed in this study if they have local mechanical stability (local jamming), positive lowest vibrational mode (collective jamming) and percolation (cluster spans the cell).  The latter has been found to be necessary \cite{Wentworth2020} as sufficiently dense pins enable the presence of mechanically-stable finite clusters. Unsupported particles, “rattlers”, are excluded from further analyses of contacts, force statistics and elastic moduli.

We use two approaches to find the jamming transition.
The volume fraction, $\phi(p)$, for these jammed configurations with different final pressures $p$ can be linearly extrapolated to $p=0$ to identify configuration-averaged, critical volume fraction $\phi_c$.   Alternatively, one could identify the $\phi$ at which a distribution of configurations has a point of inflection in its jamming probability \cite{Ohern2003,Wentworth2020}. Both criteria agree in our work. Further, for analyses which seek scaling behavior,  a pressure sweep protocol employs several hundred initial configurations and for each, modifies particle size in a search for a target pressure, which is successively reduced.

About units: Unless otherwise stated, for derived quantities like pressure and elastic moduli,  we use as an energy unit $\epsilon$ and as a length unit $r_s$.  About notation: While $N$ is used above to indicate an input parameter in the simulation, below we will sometimes refer to the number of particles in the rigid, spanning cluster at jamming. This excludes rattlers, the fraction of particles which have zero contacts at minimum energy.   As is conventional in the jamming literature, we will not introduce new notation; but will try to be clear in stating when $N$ excludes or includes rattlers.

\begin{figure}[ht]
\begin{center}
\includegraphics[width=\linewidth]{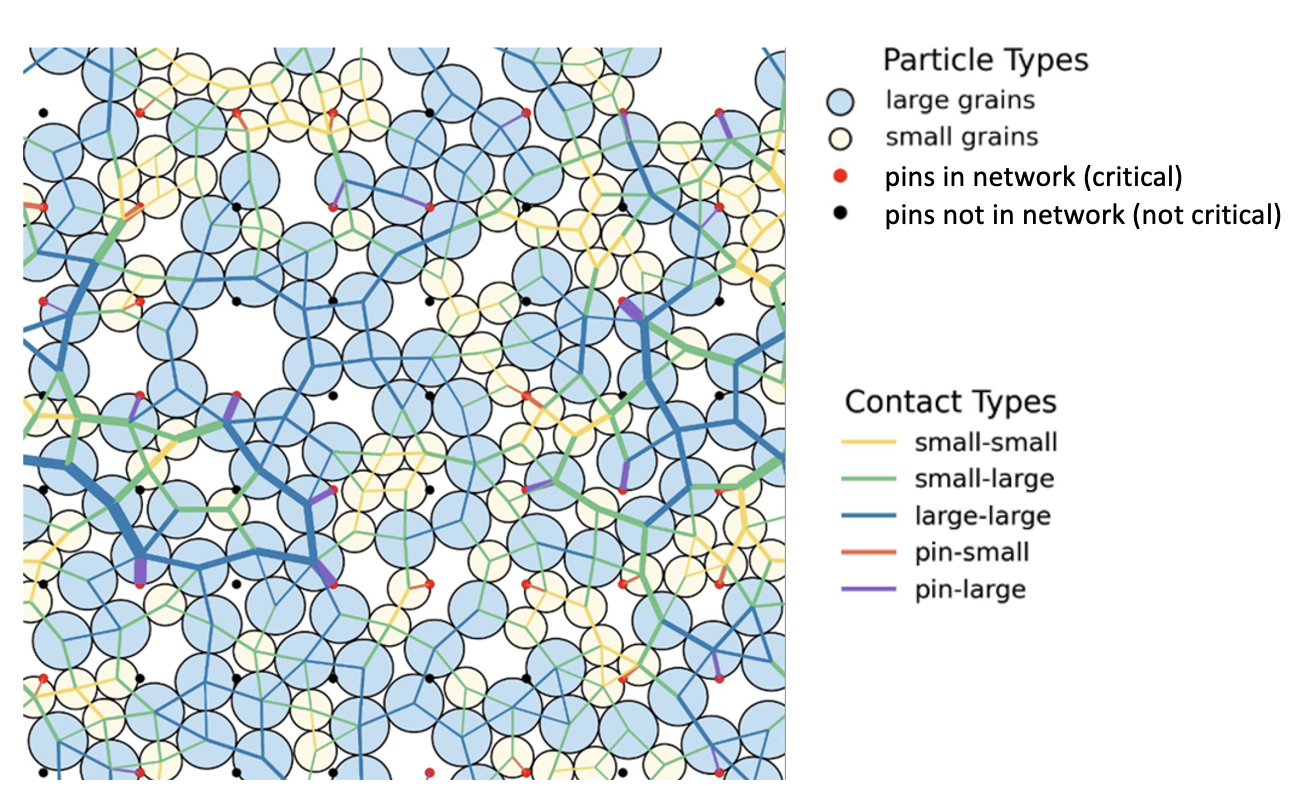}
\end{center}
\caption{Jammed configuration with $\lambda = 0.33$ and $\alpha = 0.23$ corresponding to $N_f = 64$ pins arranged in a square lattice, highlighting the network of interparticle forces. Bond colors differ by contact type and widths indicate strength of force. Pins which (fail to) contact a particle are (black) red. Pins are greatly magnified for visibility\label{fig:first}.}
\end{figure}

Figure \ref{fig:first} illustrates a sample jammed configuration of particles formed around a square lattice of 64 pins, with $\alpha = 0.23$ and $\lambda = 0.33$. At low pressures, typically only a fraction (roughly half) of pins participate in the stability of the configuration.

\section{Jamming threshold}
\begin{figure}[ht]
\begin{center}
\includegraphics[width=0.9\linewidth]{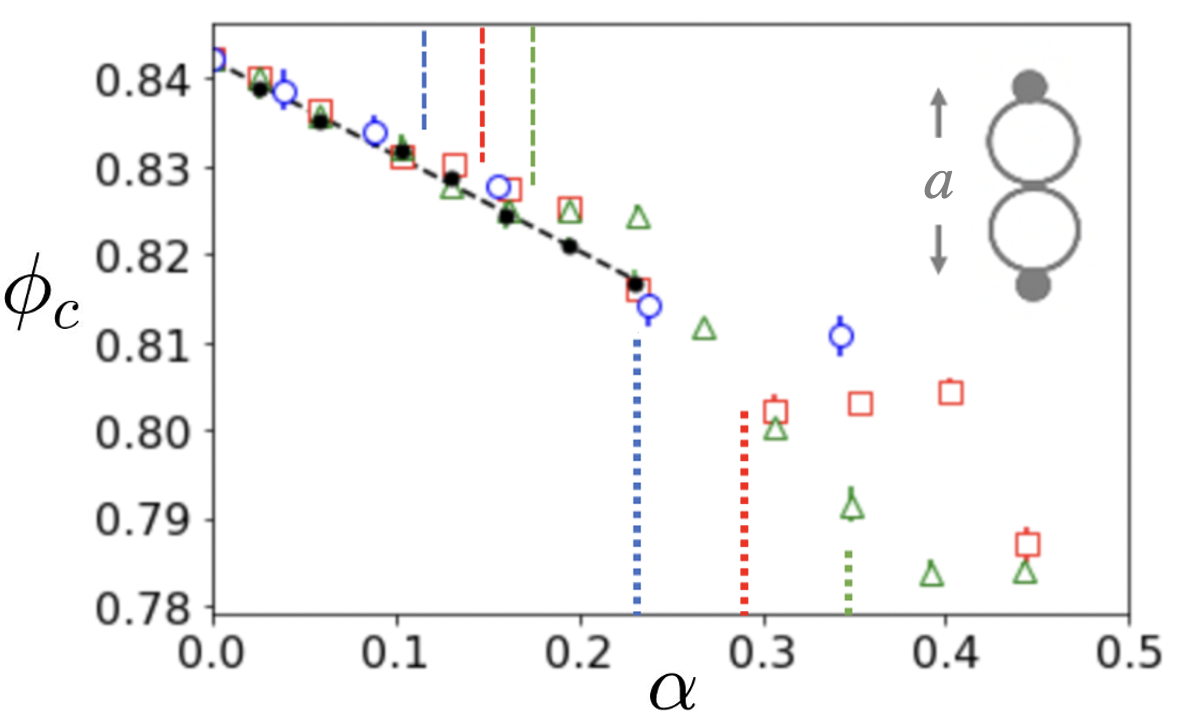}
\end{center}
\caption{Jamming packing fraction $\phi_c$ vs pin to particle ratio $\alpha$ for square (red, $\square$), triangular (green, $\triangle$), honeycomb (blue, $\circ$), and random (black, filled circles) pin lattices. 
At low $\alpha$, $\phi_c$ collapses for all lattices and is linear, with slope in rough agreement with a mean field, isostaticity argument. As $\alpha$ increases,  there is deviation from linearity and plateaus which are commensurate with values of $\lambda$ (dashed and dotted lines, $\lambda = 0.18$ and $0.25$ for large and small particles respectively) where pairs of particles in contact would first experience excluded volume from adjacent pins (inset sketch).} 
\label{fig:threshold}
\end{figure}

At low pin densities, Figure \ref{fig:threshold} aligns with results from a wealth of earlier studies \cite{Brito2013, Reichhardt2012, Graves2016, Wentworth2020,Peter2018} -  a reduction in  $\phi_c$ with increasing pinned particle concentration. Here, this is a rather trivial result since pins occupy negligible volume, and serve only to stabilize the particle network. While this work emphasizes ordered pin lattices, we observe the result that at low pin densities, when typical obstacle separations are very much greater than a particle diameter, randomly pinned lattice thresholds are in full agreement. Moreover, the linearity of these random lattice results extend to higher pin densities, allowing us to obtain a more precise estimate of the initial slope of $\phi_c(\alpha)$. 

 There are practical issues to studying very dense random pin ``lattices" for example, they tend to stabilize many finite clusters. This limit, however, could prove interesting for future work: we note that at high random pin densities, we should approach a static version of the random Lorentz gas (RLG), a model of a single particle moving among many pins which has been very useful for studying the connection between real glassy systems and mean-field theory. \cite{Manacorda2022, Hu2021, Hu2021b}. It seems likely that the RLG can be viewed in some sense as the limit of large pin density, when pins are placed randomly rather than on a lattice.  Just as the RLG has proven useful for studying corrections to mean-field theory in the dynamics, dense random pinning may prove useful to study corrections to mean-field theory in mechanical properties. 

Interestingly, Figure \ref{fig:threshold} shows that $\phi_c(\alpha)$ evolves from a lattice independent, linearly decreasing function to a more complicated form, featuring plateaus which begin at lattice-dependent values of $\alpha$.  The onset of major plateaus roughly corresponds to $\lambda = 1/4$ (indicated by dotted lines for each lattice geometry),  when two small particles can no longer ``fit" between neighboring pins.  Also visible is the departure from linearity at roughly $\lambda = 0.18 $ (dashed lines) when the same is true for two large particles between pins.  Clearly, there must exist other ``magic numbers" when small clusters in a maximally random jammed packing \cite{Torquato2015} would be disrupted by pins. The slight increase in $\phi_c$ with pin density (in square and triangular lattice plateaus) is to our knowledge the first time one has observed fixed degrees of freedom {\em raising} a jamming threshold by disrupting packing.  

The initial linear decrease of $\phi_c(\alpha)$ in Figure \ref{fig:threshold}, which continues to the highest densities for randomly-spaced pins, has been explained by the argument that pin-separation, $a$, replaces the correlation length which diverges at jamming \cite{ Reichhardt2012,Graves2016}.  One finds $\phi(0) - \phi(\alpha) = - m  \alpha$ with $m = 0.11$. The order of magnitude of this slope is surprisingly well-estimated by a simple mean field, counting argument. The number of contacts at jamming is given by the Maxwell criterion for isostaticity: critical contact number $z_c = 4$ in $d=2$ dimensions\cite{Alexander1998}. There are a fraction of $f$ pins which each provide a contact with a single particle. That is, a fraction $f$ of pins are critical to the stability of the system; whereas the remaining fraction experience zero force and can be removed with no impact on stability. (Interestingly, we find that $f$ has no observable trend with pin density, nor are adjacent critical pins spatially correlated.) Further, it is overwhelmingly unlikely that these tiny pins are in contact with more than one particle at Point J.  In the limit $\alpha \rightarrow 0$ this suggests that the critical number of particles, $N_c$ required for jamming with a fixed lattice and box size is
\begin{equation}
    N_c(\alpha) - N_c(0) = -f N_f / 4. 
\end{equation}
Substituting $f = 0.55 \pm 0.05$ from analyzed configurations, in the limit of small $\alpha \approx \frac{N_f}{N} \phi_c(0)$, and upon converting from $N_c(\alpha)$ to $\phi_c(\alpha)$, we find $m = 0.16 \pm 0.02$. One might argue that this crude counting argument overestimates the slope in the absence of particle rearrangement or deviation from the jammed structure without pins. (Four pins must be closely spaced if they are to replace contacts from missing particles.) However one expects rearrangement, which is to say deviation from the jammed structure without pins, as described below.

\section{Excess bonds, Contact numbers and Scaling behavior}
\label{sec:contacts}

Above the critical packing fraction, $N$ non-rattler particles are held in place by neighbors. The isostatic limit is achieved when the average number of contacts felt by particles is the minimum required for collective stability, when the number of particulate degrees of freedom equals the number of constraints \cite{Alexander1998, Torquato2001}.  Point J is  isostatic in simple models like polydisperse hard and soft frictionless spheres, as well as grains with circular asperities which model friction \cite{Ohern2013}, but is the exception in richer models of granular and soft matter, including anisotropic particles and other frictional models \cite{Donev2004, vanHecke2010longer}.  In the presence of pins, the system still has $N d$ degrees of freedom, but the translational zero modes are absent. 
Thus the criterion for isostaticity is generalized \cite{Brito2013, Wentworth2020} as having $N_{\mathrm{iso}}^{\mathrm{bonds}}$ interactions or ``bonds" between particles 
where
\begin{equation}
N_{\mathrm{iso}}^{\mathrm{bonds}} = dN- qd \ 
\end{equation}
with $q=1$ or $0$, without and with pins respectively. 

In order to support a finite pressure (or equivalently, have a positive bulk modulus), there must exist a set of nonzero bond compression forces that produces no net force on any particle. Such a vector of forces is known as a state of self stress, and in order to have exactly one such vector an extra contact is required \cite{Calladine1978,Goodrich2012,Lubensky2015}, giving

\begin{equation}
N_{\mathrm{min}}^{\mathrm{bonds}} = dN- qd +1\ 
\label{iso}
\end{equation}
At finite pressure, $p$, one defines $N^{\mathrm{bonds}}_{\mathrm{excess}}(p)$ as the number of bonds over and above $N_{\mathrm{min}}^{\mathrm{bonds}}$.

\begin{figure}[ht]
\begin{center}
\includegraphics[width=0.9\linewidth]{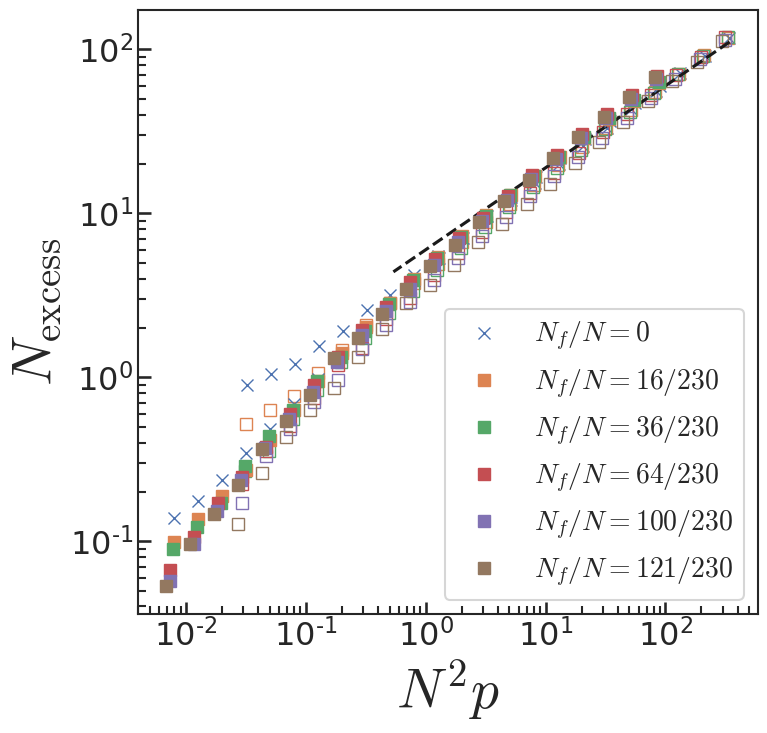}
\end{center}
\caption{Number of excess bonds as a function of $N^2 p$, a scaling which collapses finite-size effects as in ref. \cite{Goodrich2014}. Filled symbols: $N=230$, open symbols: $N=920$. Here rattlers must be excluded from $N$ to achieve collapses since the rattler fraction depends strongly on pin density.  The number of excess bonds goes to zero at zero pressure, except for a small number associated with localized states of self stress in $2d$ bidisperse systems \cite{Goodrich2014}; note that pins appear to suppress this effect. Outside of the finite-size region, $N_{\mathrm{excess}} \propto \sqrt{p}$ regardless of pin densitiy.}
\label{fig:Nexcess}
\end{figure}

We find that $N_{\mathrm{excess}}^{\mathrm{bonds}}(p) \rightarrow 0$ as $p \rightarrow 0$ in all lattice geometries studied, even when the number of pins rivals the number of particles. Figure \ref{fig:Nexcess} illustrates the relationship between $N_{\mathrm{excess}}^{\mathrm{bonds}}$ and $p$ for the square pin lattice.  One observes the expected crossover from a  low-pressure regime finite-size scaling \cite{Goodrich2014} to one which scales as $N_{\mathrm{excess}}^{\mathrm{bonds}} \sim p^{\beta}$ where $\beta = 1/2$. This well-accepted critical value which has been repeatedly observed in experiments and simulations \cite{Durian1997, Ohern2003, Katgert2010} is independent of pin density.

\begin{figure}[ht]
\begin{center}
\includegraphics[width=0.9\linewidth]{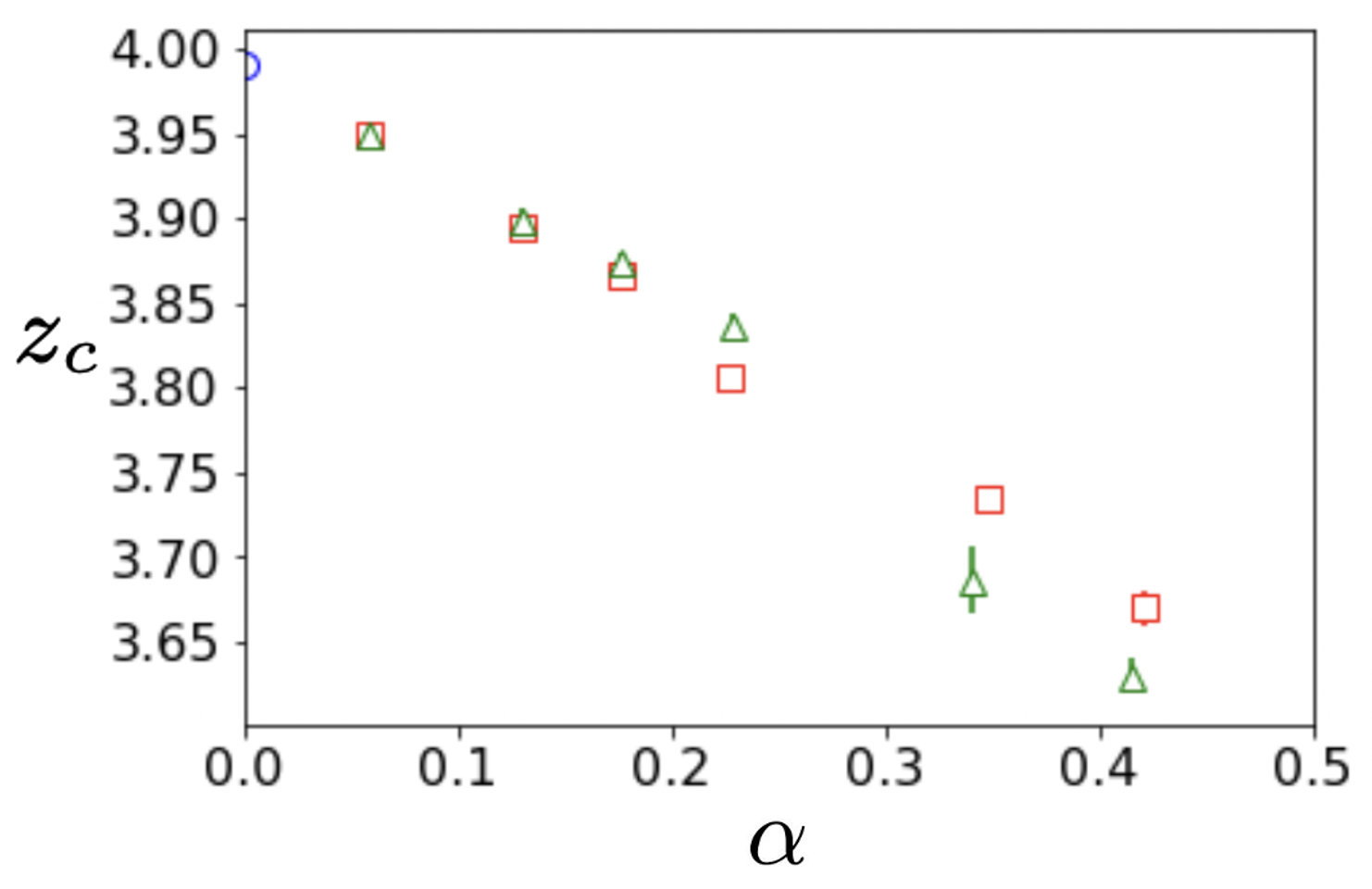}
\end{center}
\caption{The critical contact number $z_c$ vs. pin to particle density ratio $\alpha$. 
To determine $z_c(\alpha)$, we fit a power law of the form $z(p) = z_c (\alpha) + p^{\beta}$ and extrapolate to $p=0$. Colors and symbols are as in Figure \ref{fig:threshold}.}
\label{fig:Z}
\end{figure}

There is a jump from $z=0$ to $z=z_c$  as Point J is approached from $\phi < \phi_c$  \cite{vanHecke2010}. In the absence of pins,  $N_{\mathrm{excess}}^{\mathrm{bonds}}$ is trivially proportional to $N (z-z_c)$; scaling behavior is often pitched as $z - z_c \sim p^\beta$. In the presence of pins it is no longer true that  $Nz$ is twice the number of bonds,  since a particle-particle bond supports two particles, but a pin-particle bond supports only one. Thus, no simple stability argument leads to  a definition of $z_c(\alpha)$, which is free to differ from the zero pin value of $z_c(0) = 2d - 2d/N + 2/N$ \cite{Goodrich2012}. Indeed, $z_c(\alpha)$ is a decreasing function as seen in Figure \ref{fig:Z}, despite the system remaining isostatic at Point J. This reduction in contacts has been observed in previous work on square pin lattices \cite{Wentworth2020}.  Comparison with Figure \ref{fig:threshold} suggests that when $\phi_c$ is higher for a given lattice (less facilitation of jamming), $z_c$ is also higher.  This makes sense; the pin lattice permits lower coordination, hence less material in the rigid component of the structure. 
Also previously observed has been a growth in the fraction of rattlers with pin density.  A jammed system made with traditional materials, which is both less coordinated and less dense thanks to pins (in 3d, rods) has potential use as a microengineered material. Further pin-related modifications of the material, in the form of the force network and elastic moduli, are discussed below.

\section{Force networks}

\subsection{Probability distribution functions}
 Experiments using microscopy or photoelasticity provide empirical measurements of the distribution of contact forces in granular packings \cite{Amon2017, Brodu2015}. Experimental force data can also inform theoretical models which produce simulations of realistic jammed materials \cite{Zhang2005}.
In any static packing, one finds particles in mechanical equilibrium with a nontrivial distribution of forces. The fragile, jammed state has long been recognized to have a distribution of forces with extended force chains \cite{Cates1998} that lead to probability distributions with ``long tails"\cite{vanHecke2005}, when compared with solids which are either ordered, or disordered but in regimes of strong jamming, yield or flow \cite{Corwin2005, Majumdar2005}.  Exponential tails have been observed in experiments on foams, emulsions and granular packings \cite{Katgert2010, Zhang2005, Desmond2013}. Simulational models featuring a variety of repulsive forces have exponential tails \cite{ohern2001}.   One subtlety relevant to our analysis concerns the lack of self-averaging near jamming, in that the average force $\langle f\rangle$ varies substantially from configuration-to-configuration.  It has been shown that $P(F)$, where $F \equiv f/\langle f\rangle$ is normalized by each configuration's average, $\langle f\rangle$, shows a Gaussian, not an exponential tail \cite{ohern2002}. (The force ensemble, which is able to probe exceptionally rare forces, also ultimately reveals a Gaussian behavior in $d=2$ \cite{vanErd2007}.) On the other hand, if the ensemble of configurations corresponds to a single pressure, then using the ensemble-averaged mean force will also be best modeled with a Gaussian tail (something we confirmed for the data of Figure \ref{fig:contactForces} below).

Several thousand configurations at a single pressure near $p=0$ were averaged to produce the contact force distributions in Figure \ref{fig:contactForces}. 
For zero pins, the distribution is well fit by a Gaussian as expected \cite{ohern2002}. With pins, there are heavy tails which gain weight with pin density. Insets to Figure \ref{fig:contactForces} are snapshots showing chains of exceptionally strong forces which terminate on pins.  Well-known methods for robustly fitting tails of probability distributions to heavy tails \cite{Clauset2009, Alstott2014} provide fits of the form of a power law $P(F) \sim F^{-\tau}$. Comparison is also made with alternative heavy tailed distributions, which is a best practice in order to determine  goodness-of-fit \cite{Alstott2014}. Alternative forms are moderately competitive; but on the heuristic assumption that tails obey a power law the exponents 
are shown in Table \ref{table:powerlaw}. In the table,  $\tau$ is the maximum likelihood estimator for the  power \cite{Clauset2009}. The `tail" of the distribution is set by minimizing the Kolmogorov-Smirnov (KS) statistic. This procedure is sound for $N_f = 36-121$; however a KS minimum with a corresponding plateau in $\tau$ does not exist for $N_f = 0, 16$.

Apparent in Figure \ref{fig:contactForces} and Table \ref{table:powerlaw} is
that the shape of the tail is progressively less-well parametrized by $\alpha$ as pin density increases. The jamming threshold in Figure \ref{fig:threshold} supports this notion,  as well as suggesting the possibility that for sufficiently high pin densities the exponent $\tau$ is marginally better parametrized by $\lambda$, hence the pin separation in units of particle size.  

\begin{table}[h!]
\centering

\label{yourtable}
\begin{tabular}{|c|ccc|ccc|}
\hline
\multirow{1}{*}{ $N_f$} & \multicolumn{3}{c|}{Square Lattice} & \multicolumn{3}{c|}{Triangular Lattice} \\ \hline
                      & $\alpha$         & $\lambda$           & $\tau$            & $\alpha$      & $ \lambda$ & $\tau$    \\ \hline
0                     & 0              & 0              &{\em g}        & 0      & 0 & {\em g}    \\ 
16                     & .058             & .112             & -          & .058      & .104 & -    \\ 
36                     & .130             & .167             & 8.71       & .130      & .156 & 8.09    \\ 
64                     & .228             & .222             & 5.18           & .230      & .207 & 6.22      \\ 
100                    & .351             & .275             & 4.09            & .346      & .254 & 4.65     \\ 
121                    & .422             & .301             & 3.72             & .414      & .278 & 4.33     \\ \hline
\end{tabular}
\caption{Power law exponent, $\tau$, for tail of force distribution, where {\em g} indicates distribution is adequately fit as Gaussian and $-$ indicates that neither a gaussian nor power law fit is meaningful. }
\label{table:powerlaw}
\end{table}


\vspace{1cm}

\begin{figure}[ht]
\centering
\includegraphics[width=0.9\linewidth]{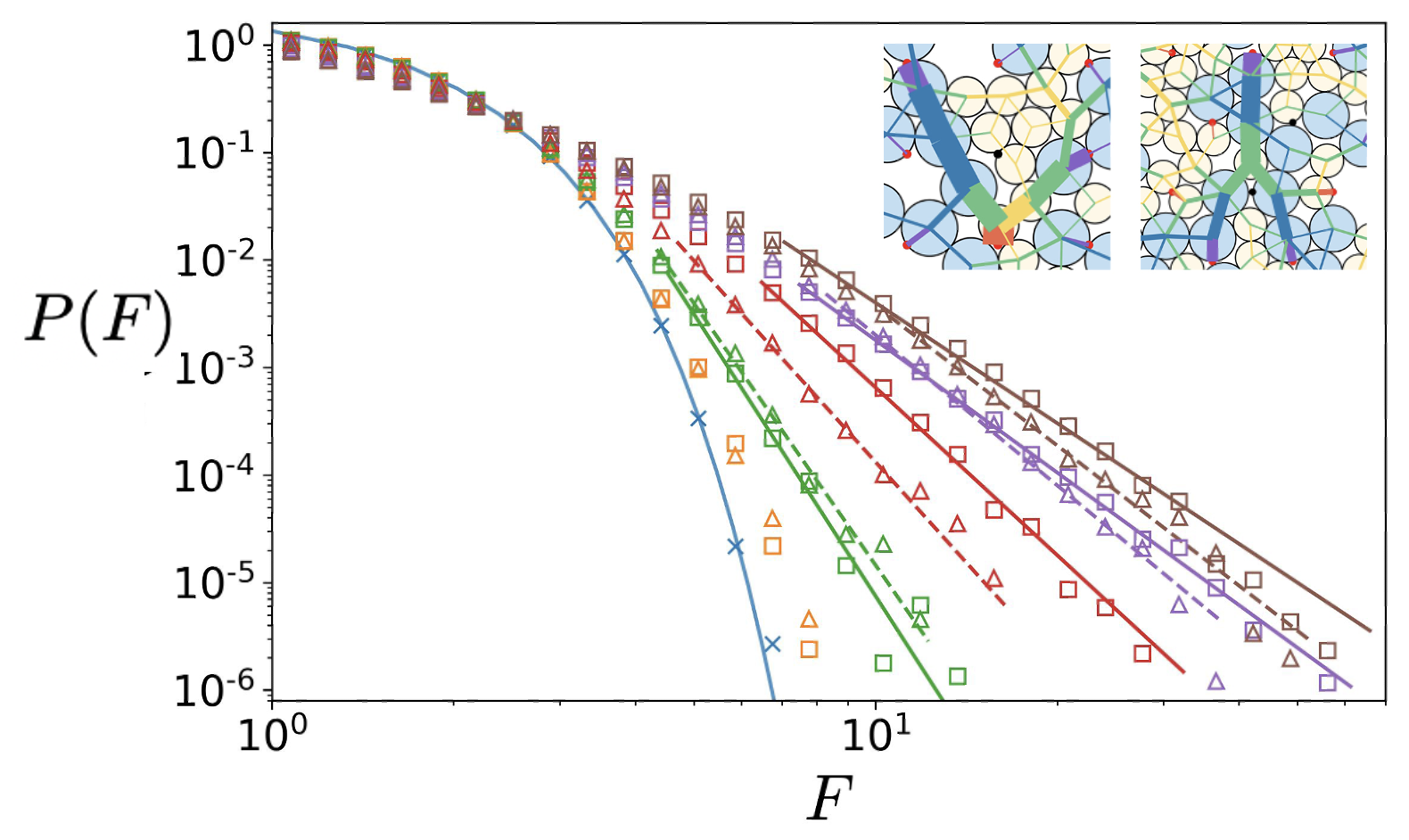}
\caption{Contact force probability distribution, $p(F)$, for square ($\square$) and triangular ($\triangle$) lattices and $N=230$ particles, where $F$ is normalized by the 
single-configuration average.  Blue, orange green, red, purple and brown correspond to 0, 16, 36, 64, 100 and 121 pins, with densities and power law exponents $\tau$ as shown in Table \ref{table:powerlaw}. 
Data at each pin density come from a single fixed (very low) pressure $p$; at different pin densities $p$ ranges from $7.5\times10^{-8}$ to $7.9\times10^{-8}$. Insets of the figure (left, square and right, triangular lattice) show strong force chains that end at pins. 
}
\label{fig:contactForces}
\end{figure}

When will a distribution of granular forces exhibit heavy and possibly power law tails? For increasingly coordinated packings, Ref. \cite{vanErd2007} shows increasingly heavy tails (albeit in the limit of forces much rarer than a non-force-ensemble study like ours can resolve). The q-model \cite{Coppersmith1995} for bead packings under gravity shows that exponential tails are the rule for a wide variety of probabilistic models for near neighbor forces, so long as force balance is required on each bead.
Mean field theory and simulation as well as exact calculation \cite{Coppersmith1996} reveal only one exceptional case. 
This critical model, in which forces are transmitted in unbranched chains through layers of beads,  yields a power law distribution of weights, $w$, on a typical bead: $p(w) \sim w^{-c}$ with $c = 4/3$ in $d=2$. While more theoretical work on a model with pins is indicated, based q-model results we might speculate that $\tau = 4/3$ is a lower limit on the scaling exponent we observe.

\begin{figure}[ht]
\begin{center}
\includegraphics[width=0.9\linewidth]{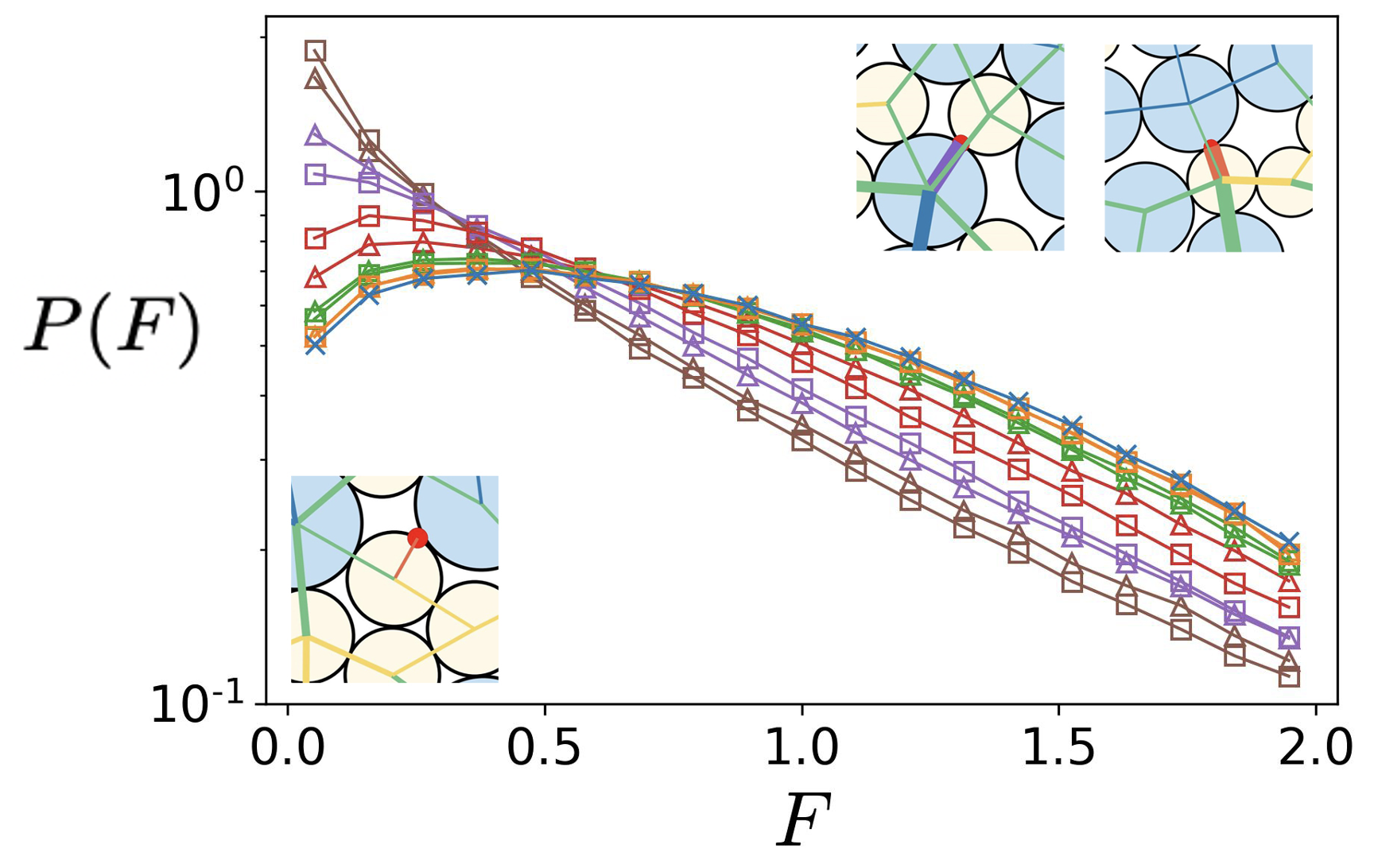}
\end{center}
\caption{Low-$F$ end of the force distribution of Figure \ref{fig:contactForces}, showing an enhancement of the likelihood of extremely weak forces, which grows with pin density.  Data at each pin density come from a single fixed (very low) pressure $p$; at different pin densities $p$ ranges from $7.5\times10^{-8}$ to $7.9\times10^{-8}$. The insets show examples of pins serving as, on the lower left, a contact for a buckler and, on the upper right, enablers which contact either one or (extremely rare at low pressure) both particles involved in the weak bond. }
\label{fig:weakForces}
\end{figure}

In the presence of pins, the distribution of forces becomes much broader.  Figure \ref{fig:weakForces} shows that not only unusually strong forces, but also unusually weak forces are much more common.  The scaling behavior of weak forces is well-studied, and there are known inequalities that relate scaling of the low-$F$ region of $P(F)$ to the distribution of gaps between particles \cite{Charbonneau2021}.  While we leave the analysis of these scaling exponents for future work, we note here that pins are effective at serving two purposes involving weak forces which come into play when there is a localized rearrangement.  One is to provide weak lateral forces to support bucklers \cite{Charbonneau2015}, particles with the minimum number of contacts. The other is to provide forces roughly collinear with the line between particle centers. We term such pins ``enablers" because, like pins which support bucklers they enable unusually small interparticle overlaps, hence forces. 
Examples of bucklers and enablers are shown in the inset of Figure \ref{fig:weakForces}.

\subsection{Locations of contacts}

\begin{figure}[ht]
\begin{center}
\includegraphics[width=\linewidth]{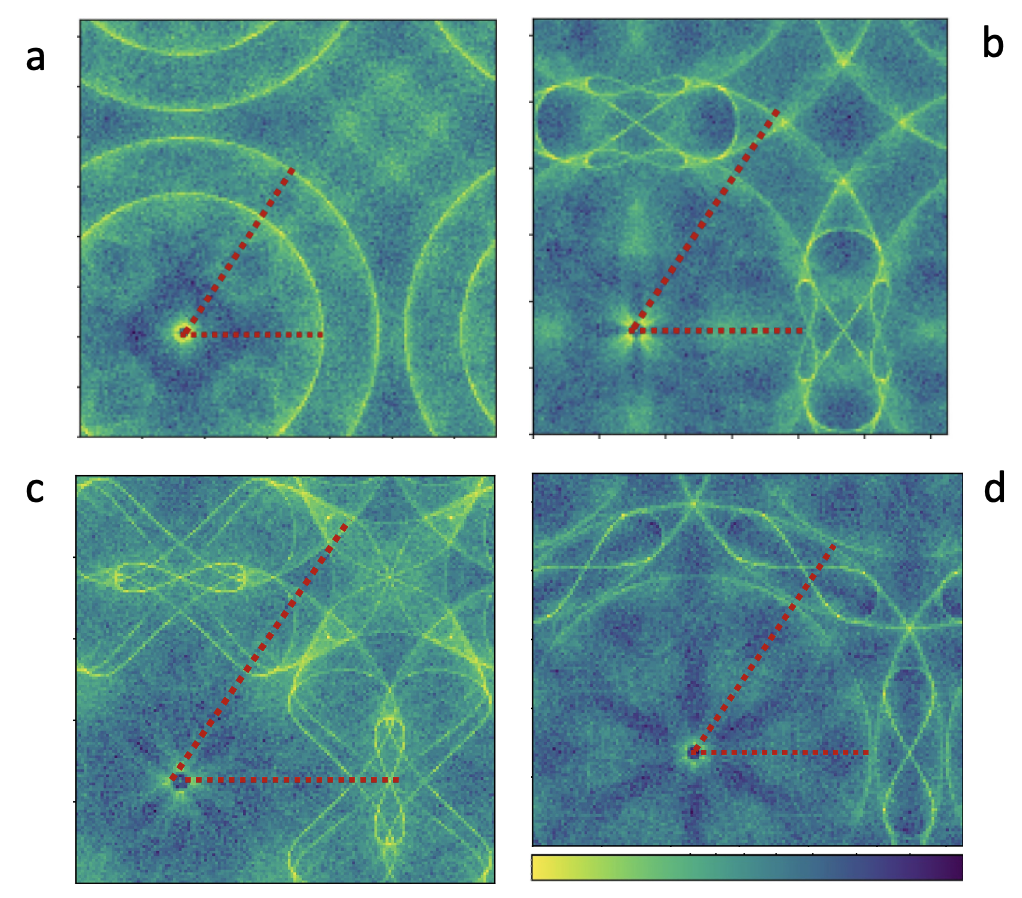}
\end{center}
\caption{Heatmaps of the likelihood of locations of bonds between particles, for $N = 230$.  Red dashed lines denote diameters of small and large particles. Correspondence between $N_f$ and particle density is as in Table \ref{table:powerlaw}. Color bar represents a logarithmic scale, from likely (yellow) to unlikely (purple). This range varies slightly between sub-figures; but all correspond roughly to one order of magnitude.   a: $N_f=36$ square.  b: $N_f=64$ square.  c: $N_f=100$ square.  d:$N_f=64$ triangular.  }
\label{fig:heatmaps}
\end{figure}

\begin{figure}[ht]
\begin{center}
\includegraphics[width=0.69\linewidth]{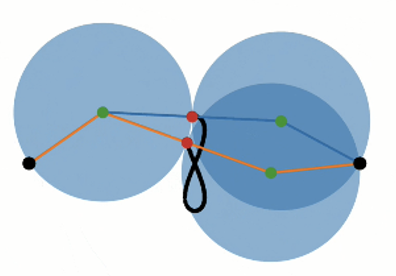}
\end{center}
\caption{If the particle on the left is imagined fixed and particle on right is imagined to move while both remain in contact with neighboring pins, the points of contact (red) trace out the figure-8 pattern shown.}
\label{fig:figure8}
\end{figure}

In our recent study on square bond lattices \cite{Wentworth2020} it was noted that pins induced spatial order. Pin-induced oscillations in the radial distribution function correlated with crystal-like peaks in the scattering function $S(\vec{k})$. Now we visualize the topology underlying these data, by mapping the likely positions of points of contact (bonds) between particles. As in the earlier study, these maps aggregate a range of pressures slightly above the jamming threshold.  Figure \ref{fig:heatmaps} depicts several cases, in which we aggregate data from statistically identical unit cells around each pin in a single configuration, and also average over several thousand configurations.   At low pin density like $N_f/N = 36/230, \ \alpha = 0.10$, the spatial frequency of contacts between particles has its maximum on circles, one particle diameter from a pin.  At a pin density where two circular loci intersect, there develop ``figure-8" interference patterns. Figure \ref{fig:figure8} shows how these come about from the constraint that two particles that pack between two pins.  Data for $N_f = 64$ has three such figure-8 features, two from small-large particle contacts, and one from large-large.  The 4-fold symmetry of the lattice is obvious here and for $N_f=100$, where there is an additional figure-8 stemming from small-small particle contacts. While aforementioned interference features are centered on the horizontal and vertical axes passing through a pin, for $N_f=100$ there are additional prominent features at $45^o$. These arise from interference of large-large particle packing between next-nearest-neighbor pins. 
Figure \ref{fig:heatmaps} additionally shows a triangular pin lattice with $N_f = 64$.  In this case there is the expected sixfold symmetry. Figure-8's arise from contacts between large-large particles; the density is just slightly below that needed for large-small contacts anchored by pins, as seen by the near touching of circular contours around adjacent pins. 

\begin{figure}[ht]
\begin{center}
\includegraphics[width=3.0in]{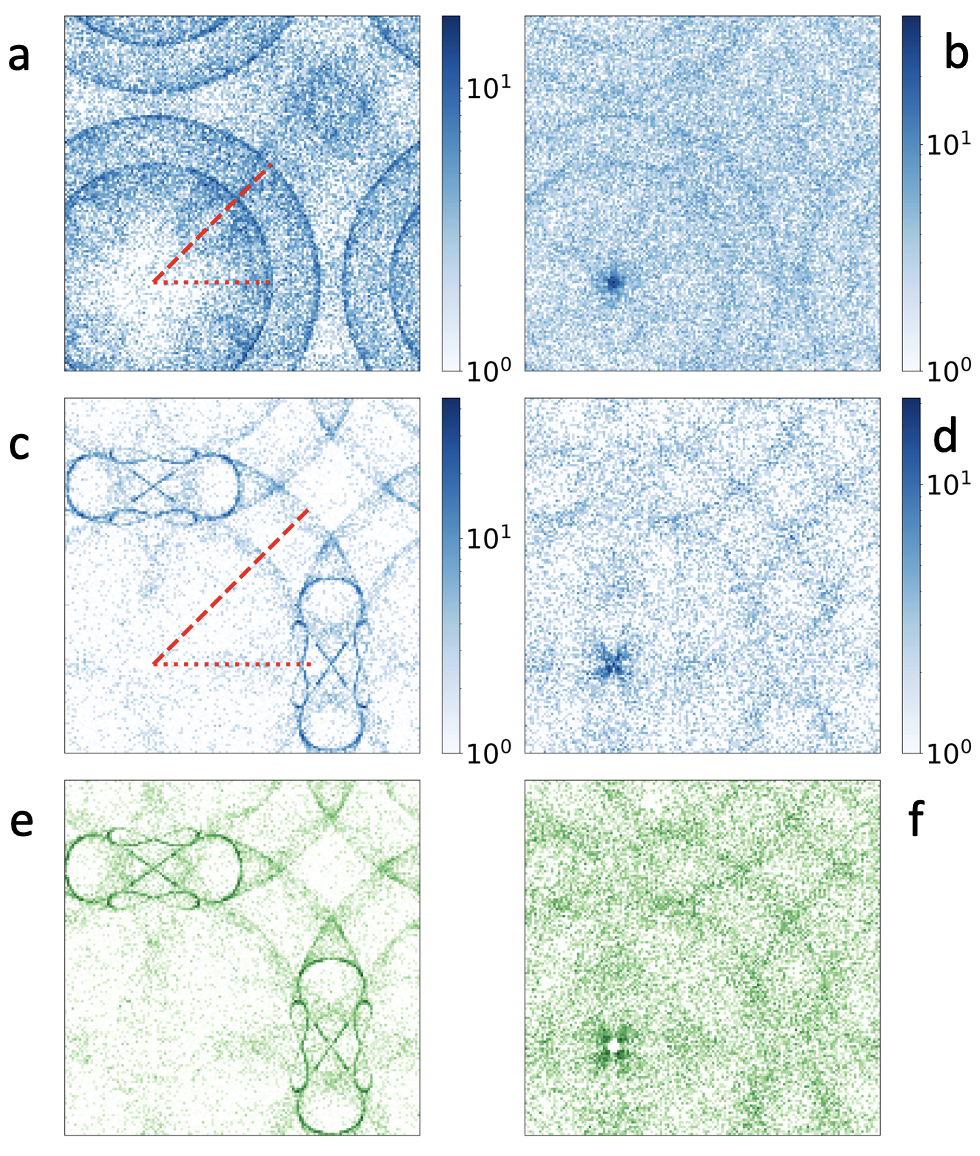}
\end{center}
\caption{Heatmaps of preferred interparticle bond locations for square pin lattices,  filtered according to forces in the top $95^{th}$ (a,c,e) and bottom $5^{th}$ (b,d,f) percentiles.   Red dashed lines denote diameters of small and large particles. All lattices are square. Correspondence between $N_f$ and particle density is as in Table \ref{table:powerlaw}. Color bar represents logarithmic scale for frequency of contact at a location. $N_f = 36$ in a, b;  $N_f = 64$ in c-f. 
a-d are for a range of pressures slightly above Point J, while e,f are for a single, very low pressure $P = 8 \times 10^{-8}$.}
\label{fig:heatmapsForces}
\end{figure}

Figure \ref{fig:heatmapsForces} shows typical bond locations as in Figure \ref{fig:heatmaps}, with bonds filtered by force magnitude.  A telling difference between these figures is that the strong interparticle bonds mapped in Figures \ref{fig:heatmapsForces}a,c,e describe particles directly in contact with a pin, or more subtly, connectioned to pins through adjacent particles. On the other hand Figure \ref{fig:heatmapsForces}b,d,f shows that the weak interparticle bonds tend to occupy locations which are quite close to a pin. The locations contributing to the strong-$F$ end of $P(F)$ of Figure \ref{fig:contactForces} are observed to change little when pressure is lowered, as seen in Figure \ref{fig:heatmapsForces}c,e. But the weak bonds become less concentrated near the pin as in Figure \ref{fig:heatmapsForces}d,f.  This might be understood by energetic considerations; the marginally stable state prefers pins in contact with a single particle.  However, when pressure is slightly increased, a second particle can move into contact with both a particle and pin, leading to a weak particle-particle contact at the low-$F$ end of the probability distribution. 

\subsection{Orientations of bonds}
\begin{figure*}[ht]
\centering
  \includegraphics[width=0.65\linewidth]{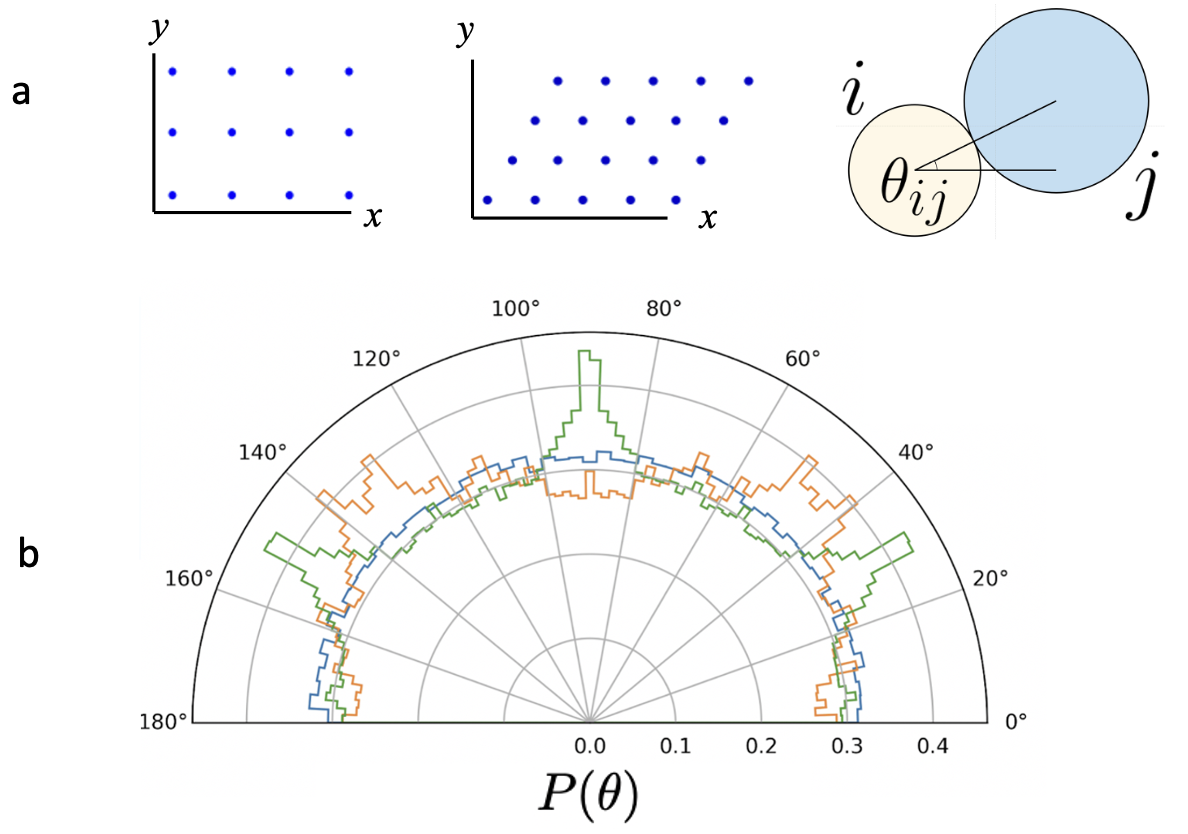}
  \caption{Histogram proportional to the probability $P(\theta)$,  that a bond makes angle $\theta$ with the horizontal axis. a: On Left and center are pin configurations indicating horizontal, $x$, direction. On right is illustrative pair of particles, $i$ and $j$, with bond angle $\theta_{ij}$. b:  $P(\theta)$ for $N = 230$. Blue: no pins, Orange: square lattice with $N_f = 100$ and $\lambda = 0.275$  Green: triangular lattice with $N_f = 64$ and $\lambda = 0.207$.}
  \label{fig:Ptheta}
\end{figure*}

In Figure 5 of Ref.~\cite{Wentworth2020}, the distribution of bond angles was compared for square pin lattices of various densities. The distribution $P(\theta)$ became progressively more anisotropic, with fourfold symmetry as one expected given a square pin lattice. In the current study, we expect to see fourfold or sixfold symmetry from square or triangular pin lattices respectively. Data shown in Figure  \ref{fig:Ptheta} from moderately dense pin lattices indeed have the expected symmetry. 

\begin{figure}[ht]
\centering
  \includegraphics[width=1.0 \linewidth]{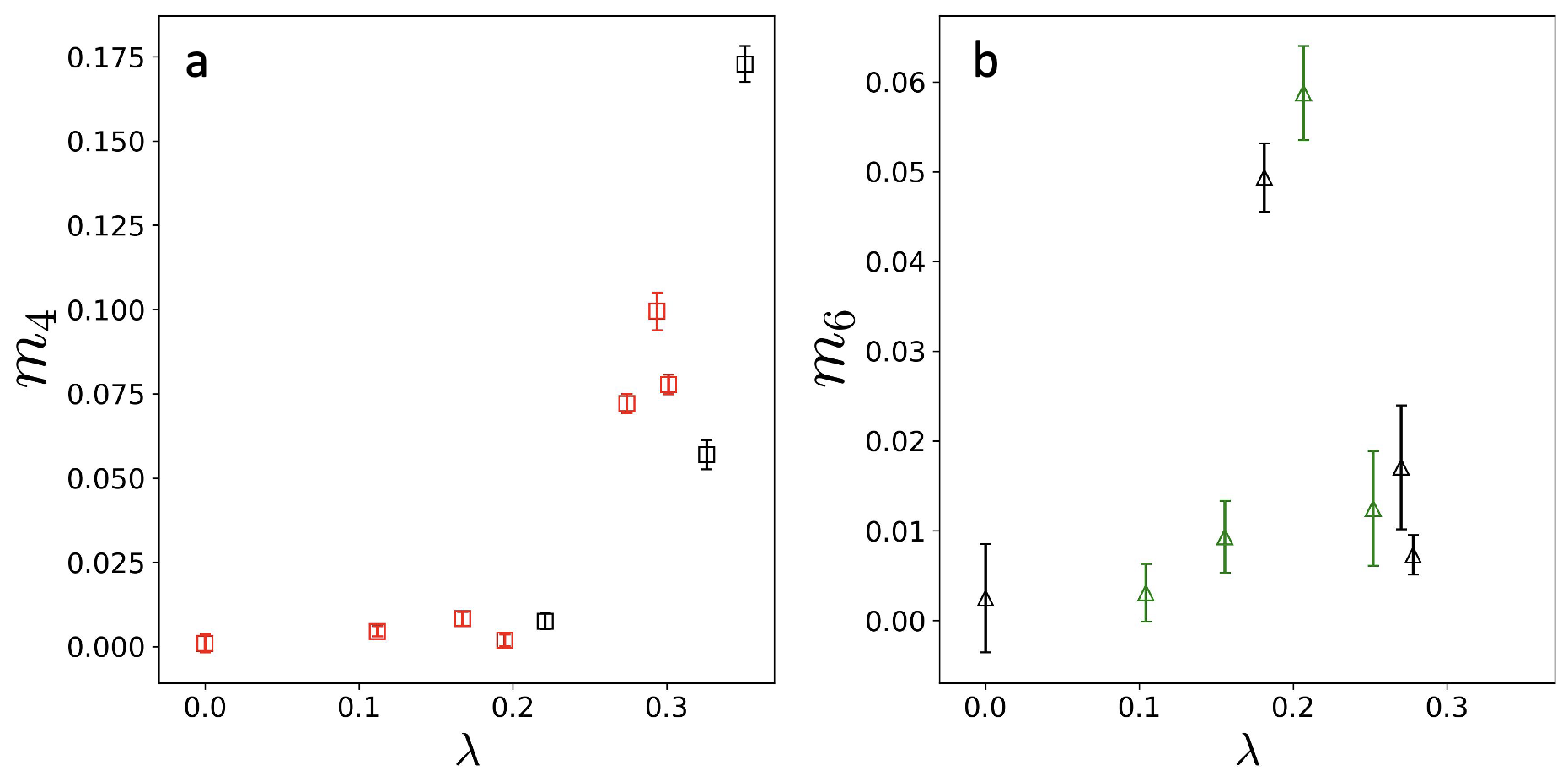}
  \caption{Order parameter, $m_q$, as a function of particle-lattice constant ratio, $\lambda$. a: Square lattice, $q=4$, b: triangular lattice, $q=6$. Color is either black or red/green depending on the sign of $  \langle Re(e^{i q \theta}) \rangle $. }  
  \label{fig:OrderParameter}
\end{figure}

There is a high level of detail in $P(\theta)$ as a function of pin density,  which is further compounded if one splits out bonds only between particles of given sizes. On the other hand, one can concisely represent the degree of angular ordering with an order parameter as in Figure \ref{fig:OrderParameter}. This shows the order parameter $  m_q  \equiv |\langle e^{i q \theta} \rangle |$ with $q = 4$ for square and $q=6$ for triangular pin lattices, and $N  = 230$ particles.  As seen in earlier work for the square lattice \cite{Wentworth2020} an angular ordering transition occurs somewhere around $\lambda = 0.25$. For the triangular lattice, this transition occurs at a lower value of $\lambda$ (hence an even lower value of $\alpha$). 
This abrupt increase in $m_q$ with cubatic or hexatic ordering is reminiscent of a phase transition as seen in the nematic order parameter in uniaxial liquid crystals with an ordering field \cite{Singh2002} or with hard rods immersed in matrices of randomly-placed hard spheres \cite{Schmidt2004}. 
Moreover, Figure \ref{fig:OrderParameter} shows that $m_q$ need not continue to increase with increased pin density above the transition. Clearly, there are certain ``magic numbers" for both square and triangular lattices, at which the bonds' orientations show evidence of being organized by pins. The onset of orientational order presents a difficult packing problem, amplified by the bidispersity of the particles. In Fig. \ref{fig:OrderParameter}, the onset distance between pins for orientational order to emerge is the length of a linear two particle bridge composed of small particles, but for the triangular lattice, is quite close to the distance spanned by a two particle bridge composed of large particles. It is quite likely that the decreasing state space for clusters of not only two, but three (and more) particles as pin density increases is a significant contributor to orientational ordering.

\begin{figure}[ht]
\centering
  \includegraphics[width=\linewidth]{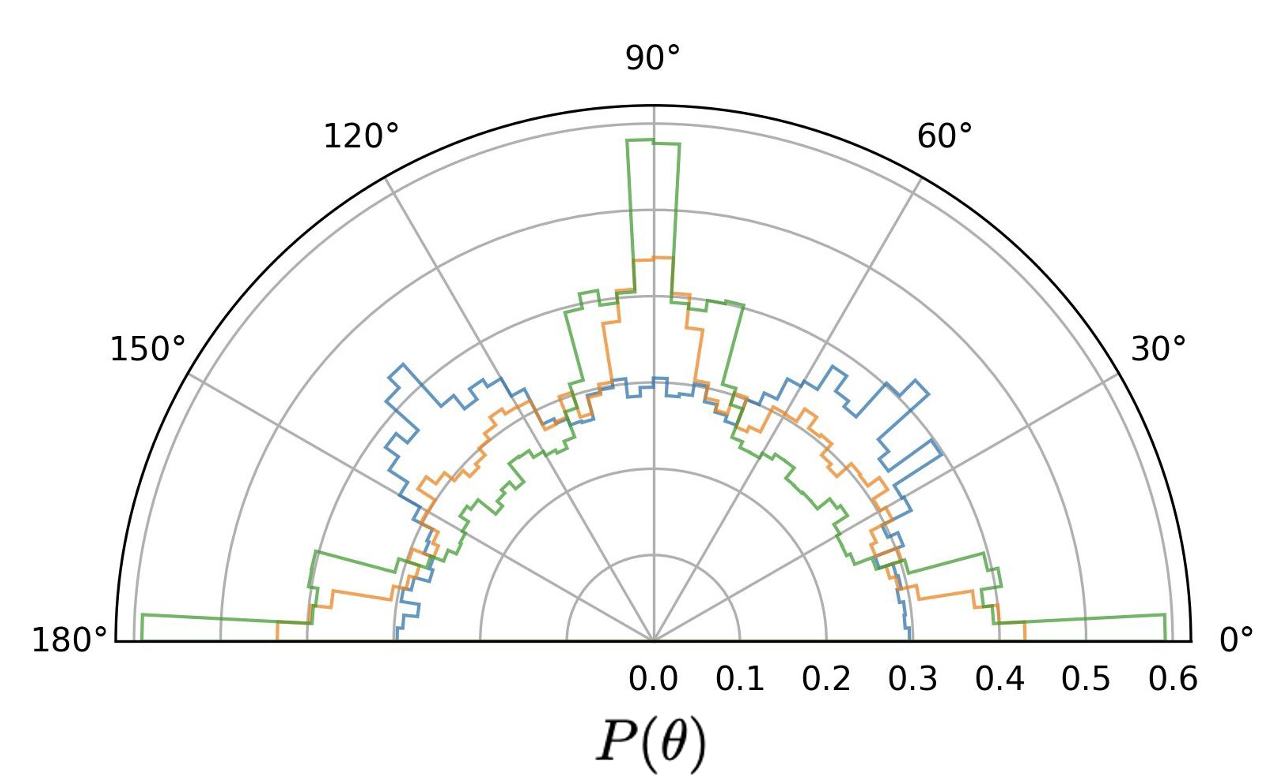}
  \caption{Histogram proportional to the probability $P(\theta)$,  that a bond makes angle $\theta$ with the horizontal axis for the square pin lattice with $N = 230$.  Blue, orange and green correspond $N_f = 121, \ 144$ and $169$, which are the three highest values of $\lambda$ analyzed in Figure \ref{fig:OrderParameter}a. }  
  \label{fig:PofThetaForOP}
\end{figure}

 Finally, while $m_q$ is a positive definite quantity as defined, we use color - black vs. red or green - to indicate changes in sign in the quantity $\langle Re(e^{i q \theta}) \rangle$ .  This is an indicator of a shift in the particular way that the lattice symmetry is manifested. In particular, Figure \ref{fig:PofThetaForOP} shows such a phase shift for the square lattice; in lieu of  $\pi/4$ and $3 \pi/4$ being the most likely bond angles for $N_f = 121$, bonds are most likely oriented at $0$ or $\pi/2$ for $N_f = 144, \ 169$. This shift in most probable orientation turns out to be consequential for the elastic properties of the jammed solid.  In Figure \ref{fig:shear} we will see excellent correlation between this shift in symmetry and the behavior of the Zener ratio, which indicates breaking of the isotropy of the elastic tensor.  

\subsection{Topological analysis}
In order to better understand the multiscale character of the stress-bearing structures in the system, we leverage topological data analysis to analyze the network formed by particles (nodes of the network) and the interparticle forces (edges of the network).  Specifically, we follow the procedure presented in Ref.~\cite{MischaikowKondic} to determine the persistence of features in the contact force network as we apply increasingly aggressive filtration of the network by omitting forces below some threshold, and study how the persistence depends on both pin density and pin arrangement. The quantities of interest to emerge from this analysis are the zeroth and first Betti numbers, $\beta_0$ and $\beta_1$. For a given filtration of the contact force network, $\beta_0$ is the number of disjoint (not connected by any edges) components of the network.  $\beta_1$ counts closed paths of 4 or more edges which bound an empty region of the filtered network. These quantities have proven to correlate with the mechanical response of granular systems in useful ways. For example, the time evolution of persistent features has been shown to correlate with material failure~\cite{SMPH} and impact dynamics~\cite{PersImpact}. Of more relevance to the present study, the persistent features of packings have illuminated structural effects of particle shape~\cite{TopShape}, exposing the sensitivity of bulk properties of packings to particle-scale constraints imposed by grain geometry. Thus persistent homology is a promising strategy to explore the effect of the quasi-local constraints imposed by pins on a packing inside a pin lattice.

\begin{figure}[ht]
\begin{center}
\includegraphics[width=\linewidth]{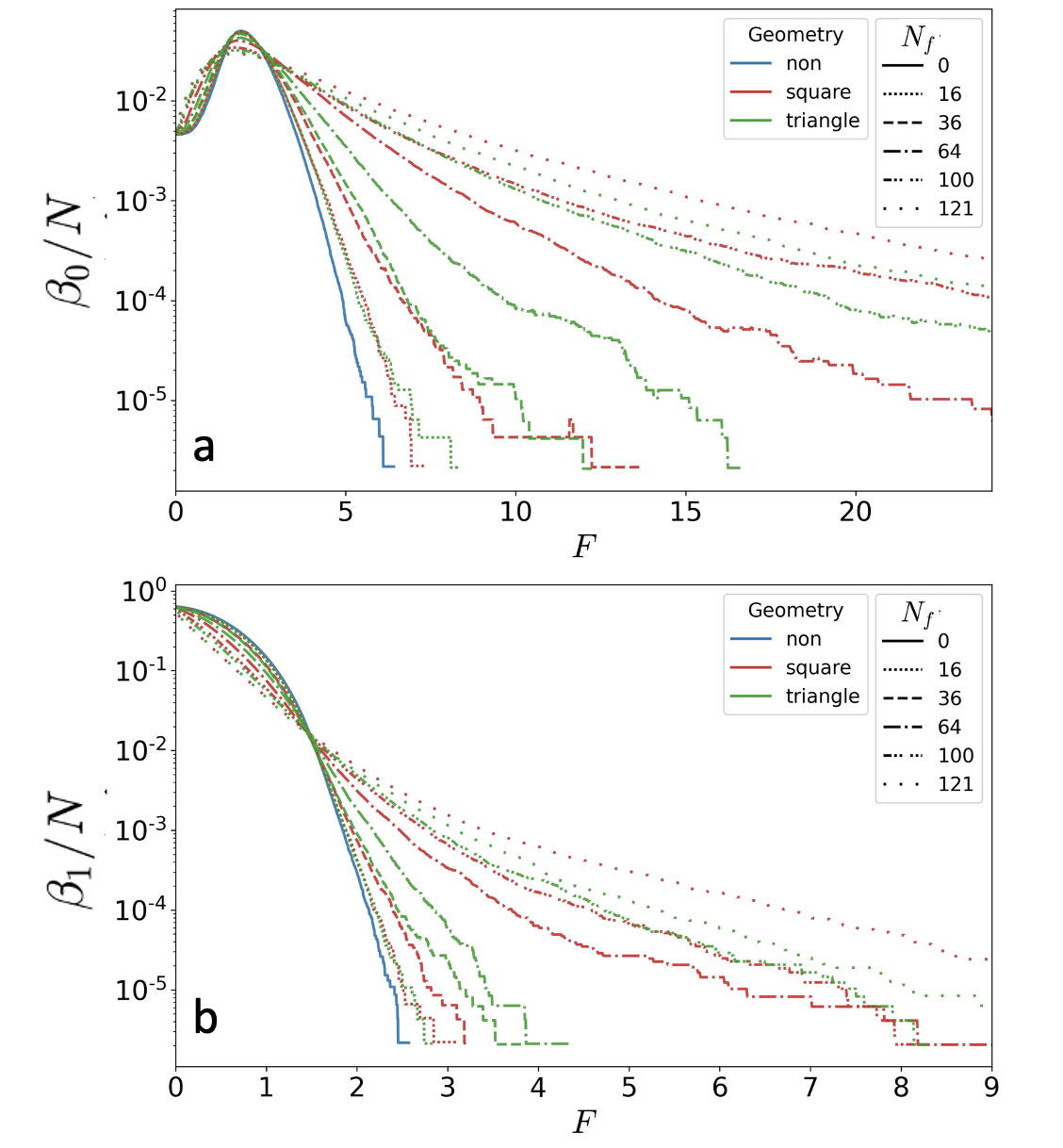}
\end{center}
\caption{ a: The zeroth and b: first Betti numbers, normalized by the number of non-rattler particles, plotted against the filtering force (normalized by the mean force). The zeroth Betti number is a measure of the number of disjoint components of the network whereas the first Betti number is a measure of the number of loops in the network.}
\label{fig:PersistentHomology}
\end{figure}

We find that the presence of pins has a dramatic impact on the topoological properties of the contact force network. Betti numbers are plotted vs filtration in Fig.~\ref{fig:PersistentHomology}. We see that, in systems with no-to-few pins, the zeroth Betti number is insensitive to very small filtrations before turning up to reach an approximately gaussian peak.  As the number of pins is increased, the small-force plateau vanishes and the peak in $\beta_0$ becomes dramatically heavy-tailed.  These basic trends appear to be qualitatively insensitive to the geometry of the pin array.  Similarly, as the pin number is increased, $\beta_1$ transitions from roughly Gaussian to an extremely heavy-tailed curve, qualitatively insensitive to the pin array geometry.

These results suggest the pins play an important role (reflected by the heavy tail in $\beta_0$) as sinks for stresses in the packings.  Chains of strong interparticle forces are able to terminate at the pins, akin to container boundaries in the Jansen effect but on a much more local scale.
We might hypothesize based on this interpretation alone, that pressures in the packing would be less correlated on scales greater than the pin spacing than for packings without pins.
Though testing this prediction is beyond the scope of the present study, the scaling of $\beta_1$ provides a complementary picture. The persistence of loop-like features at high-filtration suggests that large stresses are transmitted around regions of relatively low-stress, visibly anchored at a subset of pins.

\section{Elastic properties}

How does the presence of pins affect the response of a packing to perturbations? For a solid, the simplest description of the macroscopic response is linear elasticity, and at the ordinary jamming transition the elastic moduli show critical scalings near the jamming transition, with bulk modulus

\begin{equation}
    B \sim p^0 
\label{eq:bulkModulus}
    \end{equation}
    
    and shear modulus 
    \begin{equation}
    G \sim \sqrt{p}
\label{eq:shearModulus}
\end{equation}
in an infinite system.  In a finite system, $G$ shows clear finite-size corrections, crossing over to a plateau $G \sim 1/N$ at small values of $p N^2$ \cite{Goodrich2014}.

The elastic modulus tensor, from which these elastic constants are derived, may be defined as

\begin{equation}
    C_{ijkl} \equiv \frac{\partial^2 E}{\partial \epsilon_{ij}\partial \epsilon_{kl}}.
\end{equation}

In a normal packing, the meaning of the global $\epsilon_{ij}$ is unambiguous, and (for example) a shear $\epsilon_{ij}$ may be realized by transforming the vectors $a_j$ defining the periodic cell as $a'_i =\epsilon_{ij} a_j$.

In our packings with pins, there is no longer a unique choice of global deformation.  One may imagine deforming a sample while the pins remain fixed in their original lattice. This deformation is incompatible with the periodic simulation as it changes the ratio between the sample volume and the pin lattice unit cell volume; thus, simulating this deformation is somewhat complicated. 

A different definition of strain which is much simpler to analyze is a strain in which the array of pins is also forced to deform along with the periodic cell.  This second definition, which we adopt, also has the convenient feature of preserving the connection between the bulk modulus, the packing fraction, and the pressure which exists in pin-free packings:

\begin{equation}
    B = \phi \frac{d p}{d \phi} = \frac{\phi^2}{V} \frac{d^2 E}{d \phi^2}.
\end{equation}

In practice we calculate elastic moduli using exact linear response by inversion of the dynamical matrix.

\subsection{The shear modulus in the presence of pins}

As an ensemble, ordinary jammed packings are isotropic. Packings formed in the presence of a square lattice of pins, however, lose this isotropy, as is evident from Figures \ref{fig:Ptheta}--\ref{fig:PofThetaForOP}, and thus require an extra elastic constant to describe them completely in two spatial dimensions:

\begin{align}
B &=\frac{1}{4} \left(C_{xxxx} + C_{yyyy} + 2 C_{xxyy}\right)\\
G_{1} &\equiv C_{xyxy}\\
G_{2} &\equiv \frac{1}{4}\left(C_{xxxx} + C_{yyyy} - 2 C_{xxyy}\right).
\end{align}

We may equivalently describe the shear response by the combination of the angle-averaged shear modulus

\begin{align}
\bar{G} &= \frac{1}{2}\left(G_{1} + G_{2}\right),
\end{align}

and the Zener ratio

\begin{align}
a_r &= G_{1}/G_{2}.
\end{align}

\begin{figure*}[ht]
\begin{center}
\includegraphics[width=0.75\linewidth]{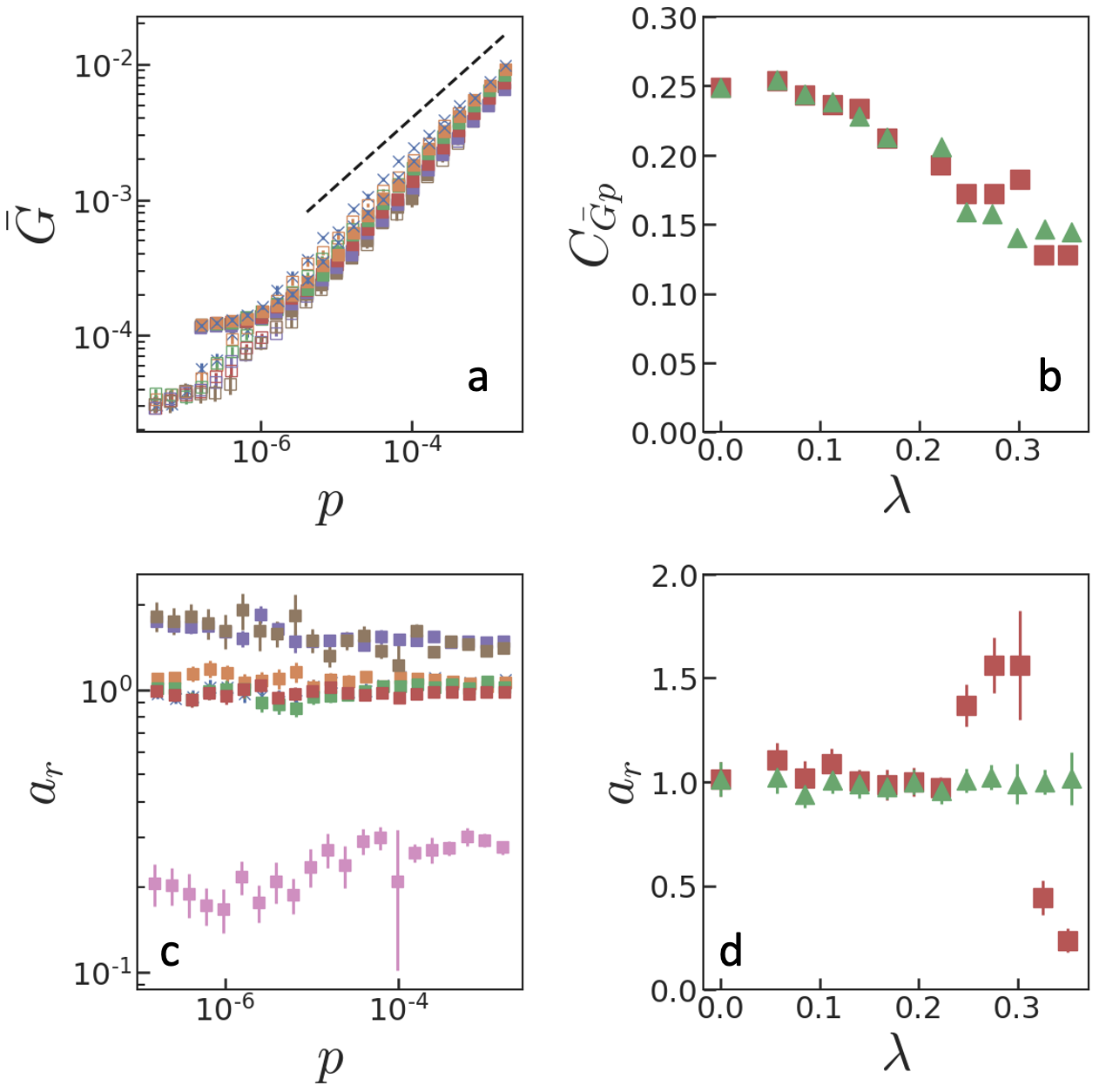}
\end{center}
\caption{Shear moduli of packings formed in the presence of pins.  a: The angle-averaged shear modulus $\bar{G}$ shows pin-density-independent scaling behaviour, with a finite-size plateau that is independent of pin density, but a different scaling prefactor. Colors are same as in Figure \ref{fig:Nexcess}(while pink represents a further higher pin density $N_f/N = 169/230$); open symbols are $N=920$ while filled symbols are $N = 230$. b: Scaling prefactor in $\bar{G} = C_{\bar{G}p} \sqrt{p}$ decreases as a function of pin density in square lattices. c:  Zener ratio, measuring cubic anisotropy in the shear moduli, is roughly, but not perfectly, independent of pressure at all pin densities for square lattices. d: For triangular lattices, the pressure-averaged Zener ratio is independent of pin density. For square lattices, a strong cubic anisotropy in elastic constants develops at the same $\lambda$ as bond-orientational order (Figure \ref{fig:OrderParameter}).}
\label{fig:shear}
\end{figure*}

Figure \ref{fig:shear}a shows the angle-averaged shear modulus $\bar{G}$ as a function of pressure for square lattices. The scaling behaviour, with the expected exponent of $1/2$ as in Eq. \ref{eq:shearModulus}, appears to be independent of pin density, and the magnitude of the finite-size plateau of $\bar{G}$ appears to be unaffected by pin density. The prefactor $C_{\bar{G} p}$ between $\bar{G}$ and  $\sqrt{p}$ outside the finite-size regime, however, appears to decrease with increasing pin density, roughly independently of pin geometry even for large $\lambda$, as illustrated in \ref{fig:shear}b.

Figure \ref{fig:shear}c  shows that the Zener ratio $a_r$ has a fairly weak pressure dependence, and thus in Figure \ref{fig:shear}d we plot its mean over all pressures as a function of pin density $\lambda$, for both square and triangular lattices. As might be expected for systems with perfect sixfold symmetry\cite{LandauLifshitz},  the triangular pin lattice induces no sign of cubic anisotropy at any pin density, while the square lattice does at sufficiently high density.  Comparing to Figure \ref{fig:OrderParameter}, we see that the  anisotropy $a_r - 1 \neq 0$ of the elastic constants coincides with the onset of orientational order $m_4$. Furthermore, when $m_4$ changes sign, the sign of the anisotropy changes as well.

One can understand this qualitatively by considering the affine part of the elastic modulus tensor. Recall that the affine moduli e.g. $G_{A}$ represent the energy cost of forcing the particles to follow the applied deformation affinely; this represents an overestimate of the true elastic moduli because this affine deformation will tend to produce net forces on each particle which must be relaxed away to reach the true sheared state (and in fact, for the shear modulus, $G_{A}$ is a gross overestimate, since it does not go to zero as $p\to 0$).   

We may easily show that

\begin{equation}
    \left(a_r\right)_{A} = \frac{\langle \sin^2{\left(2 \theta \right)}\rangle}{\langle \cos^2{\left(2 \theta \right)} \rangle} = \frac{1 - \langle \cos(4 \theta) \rangle)}{1 + \langle \cos(4\theta)\rangle} \approx 1 - 2 \mathrm{Re}\, \langle e^{i 4 \theta}\rangle.
\end{equation}

Thus, the correlation we've previously seen between $a_r$ and $m_4$ qualitatively matches that which is expected for the affine moduli. The agreement is not quantitative, however; comparison of the numerical values shows that this prediction from $m_4$ underestimates the degree of anisotropy by a factor of 2--3.

\subsection{Dependence of the low-pressure bulk modulus on pin density}

\begin{figure}[ht]
\begin{center}
\includegraphics[width=\linewidth]{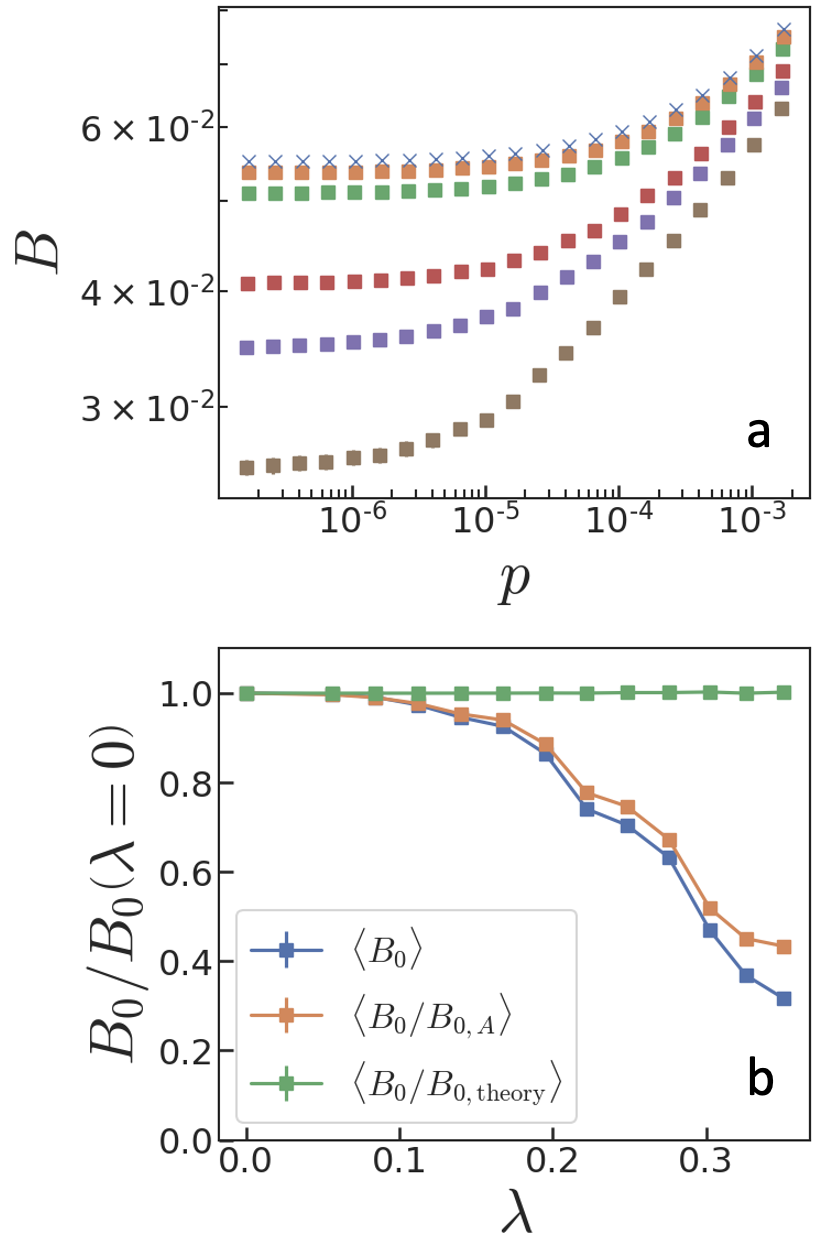}
\end{center}
\caption{a: Bulk modulus as a function of pressure for square pin lattices, showing plateau at low pressure $p$.  Colors as in Figure \ref{fig:Nexcess}; the modulus decreases at high pin densities.  b: Low-pressure bulk modulus $B_0$, divided by its value with $\lambda=0$ (no pins), in blue.  $B_0$ decreases as the pin density is increased.  Orange: $\langle B_0 / B_{0, A}\rangle$ shows a slightly weaker decrease, showing that the reduction in affine bulk modulus associated with reduced jamming density plays some role in this reduction of $B_0$, but only a small one.  Green: $\langle B_0 / B_{0, \mathrm{theory}}\rangle$, where $B_{0, \mathrm{theory}}$ is given by Eq. \ref{eq:bulk_theory}, showing perfect agreement with theory.
\label{fig:bulk}}
\end{figure}

As pressure is decreased toward zero, the bulk modulus plateaus, as seen in Figure \ref{fig:bulk}a. Thus an important result is that, as with the shear modulus,  pins do not alter a critical scaling exponent; in this case $0$, for $B(p)$ as in Eq. \ref{eq:bulkModulus}.  Yet, the plateau value, $B_0$ has a systematic dependence on pin density. 
Naively, one might assume that the introduction of pins should stiffen the system, and increase $B$.  Indeed, at fixed $\phi$, this should be true. However, as we have seen earlier, increasing pin density allows the packing to jam at a lower $\phi_c$, with larger numbers of rattlers so the rigid network is even less dense than one might expect from a knowledge of $\phi_c$ alone.  Thus the limiting value $B_0$ need not increase with pin density; and indeed the trend in Figure \ref{fig:bulk} is that it decreases.

How can we rationalize this trend? The basic idea is that introduction of pins allows lower-density, more-fragile packings to form. Qualitatively, a reduction in bulk modulus is seen in experiments on granular packings where increased polydispersity leads to increased effective porosity and reduced stiffness\cite{Petit2017}. Further, for a broad range of systems with structural disorder, increasing disorder causes lattices near isostaticity to stiffen\cite{Zaccone2011}; the flip side is that the ordering provided by pins, would have the opposite effect. 
To make this idea precise, we note that the plateau bulk modulus may be expressed  as \cite{Sartor2021}:

\begin{equation}
    B_{0,\mathrm{theory}} \approx \frac{ \epsilon N^{\mathrm{bonds}}_{\mathrm{min}}}{d^2 V} \frac{\langle \sigma f \rangle_{\mathrm{bonds}}^2}{\langle \sigma^2 f^2 \rangle_{\mathrm{bonds}}}, \label{eq:bulk_theory}
\end{equation}

where $\sigma$ is the sum of radii for a particular bond, $d=2$ in our case, and the volume $V$ is thus the area of the system. 
The left-hand factor expresses the fact that a less-dense packing has a lower \textit{affine} bulk modulus $B_{0,A}$---with fewer particles participating in the packing, fewer bonds are present and thus the energy cost of an affine deformation is reduced.  The right-hand factor accounts for the non-affine relaxation in the limit of small pressure; a packing with a broader distribution of forces is able to relax a greater fraction of the stress initially imposed by an affine compression.

We thus see that the packings with pins are expected to show a reduction in zero-$p$ bulk modulus due to both factors: a reduced $\phi_c$ and a broader distribution of forces. Figure \ref{fig:bulk}b confirms this explanation by plotting the average value of the low-pressure bulk modulus $B_0$, divided by its value at $\lambda = 0$. The blue curve shows the decay of $B_0$ as the pin density increases. Further curves show the average values of $B_0 / B_{0,A}$ and $B_0 / B_{0,\mathrm{theory}}$. The final curve is perfectly flat, showing that the theory explains the decrease of plateau bulk modulus with pin density. Dividing by the affine modulus, on the other hand, only removes a small part of the decrease with pin density. Thus, although the reduced density at jamming explains part of the reduction in $B_0$, the dominant effect is the change in non-affine relaxation, which may be related to the broadening of the distribution of forces.

We have thus found two nontrivial effects of the pin lattice on the elasticity of the resulting packings.  In the shear moduli, a cubic lattice of pins induces a cubic anisotropy, which is closely correlated with the development of bond-orientiational order at sufficiently large pin densities. In the bulk modulus, we find that the lattice of pins stabilizes packings with much lower bulk moduli than normal packings, and that this reduced modulus is intimately connected to the broader distribution of forces which is found in the presence of pins.

\section{Conclusions}
Simulated jamming of bidisperse soft discs with the addition of a lattice of fixed pins permits one to tune the jamming threshold and structure of the force network, as well as modify the solid's elastic properties.  A fragile, marginally-stable jammed solid exists at zero pressure, with the familiar value of critical exponent $\beta = 1/2$ for scaling of the number of bonds above the isostatic limit.  Also familiar are the critical exponents that describe the bulk and average shear modulus, as a function of pressure. However, both the jamming threshold $\phi_c$ and the critical contact number $z_c$ are dependent on lattice density and identity, with  $\phi_c $ showing  plateau-like features where lattice-specific, local packing effects are important. For small pin densities, investigated quantities are independent of lattice geometry. Maps of contact locations within one unit cell of the lattice reveal an inhomogeneous, patterned structure indicative of packing around pins. At sufficiently high pin density, the pattern's symmetry matches that of the lattice.   Distributions of bond angles show anisotropy; with an angular order parameter that rises from zero when particle diameter becomes comparable to pin separation.  The bond angular probability distribution undergoes detailed changes as pin density changes. For a square (but not triangular) pin lattice, this correlates with deviations of the Zener anisotropy, $a_r$, from unity. The result is a disordered solid with two distinct shear moduli.

Dramatic pin-mediated changes occur in the distribution of forces, including  enhanced probability at both weak and strong forces.  A heuristic power law model for the strong tails is employed: $P(F) \sim F^{-\tau}$, with exponent $\tau$ found to decrease with increasing pins. Considerations of persistent homology are employed, which show equally dramatic effects of pins in enhancing loop-free topological structures at low force filtration, and supporting both types of topological structure at higher force filtrations.

This wealth of structural changes has consequences for elastic behavior. In broad terms, both bulk and shear moduli decrease with pin density. A more detailed theoretical treatment demonstrates that changes in contact number, density, and enhanced breadth of force distribution all contribute to the reduction in bulk modulus, with the increased non-affine relaxation associated with the broad force distribution being the main factor. Thus, the placement of supporting pins prior to jamming is a promising technique to engineer a disordered, jammed material in two dimensions which is less dense and less rigid, with mechanical anisotropy in the form of two different shear moduli. Photoelastic experiments on grains jammed in the presence of pins, as well as extensions to other situations of interest like jamming and yielding in the presence of shear or three dimensional materials supported by fixed rodlike structures, are promising avenues for future work. 

\section{Acknowledgments}
We are grateful for helpful conversations and technical assistance from the following individuals: M. Dijkstra, A.J. Liu, B. Jenicke, A.R. Moser, L. Packer, C.J.O. Reichhardt, T.A. Witten and A. Zaccone. This material is based upon work supported by the National Science Foundation under Grant Numbers DMR-1905474, DMR-1905737, and DMR-2046551.  We further thank Swarthmore College’s Provost, Division of Natural Sciences, and Individual Donors. S. A. Ridout has been supported by PGS-D fellowship from NSERC and the Simons Foundation Cracking the Glass Problem Collaboration award No. 454945 to Andrea J. Liu.

\bibliography{References}

\end{document}